\documentclass{Interspeech}

\interspeechcameraready

\usepackage{cite}
\usepackage{amsmath,amssymb,amsfonts}
\usepackage{graphicx}
\usepackage{textcomp}
\usepackage{xcolor}

\usepackage{multirow}
\usepackage{graphicx}
\usepackage{amsmath}
\usepackage{amssymb}
\usepackage{amsxtra}
\usepackage{threeparttable}
\usepackage{subcaption}  

\usepackage{amsmath}
\usepackage{algpseudocode}
\usepackage[whole]{bxcjkjatype}
\usepackage{subcaption}

\title{RAG-Boost: Retrieval-Augmented Generation Enhanced LLM-based Speech Recognition}

\author[affiliation={1}]{Pengcheng}{Wang}
\author[affiliation={1}]{Sheng}{Li}
\author[affiliation={1}]{Takahiro}{Shinozaki}

\affiliation{School of Engineering}{Institute of Science Tokyo}{Japan}
\email{wang.p.6102@m.isct.ac.jp, sheng.li@ieee.org, shinot@ict.e.titech.ac.jp}
\keywords{Large language model, Speech recognition, Retrieval-augmented generation}

\usepackage{comment}

\begin{document}

\maketitle

\begin{abstract}
In this paper, we propose RAG-Boost (ST-ShinozakiLab Task I system), which enhances the baseline LLM-based ASR system of the MLC-SLM Challenge (task I) with a retrieval-augmented generation (RAG) module on the fly. Each partial ASR hypothesis queries a vector store of audio–text pairs and domain terms, and the retrieved results are fused with the live ASR hypotheses to fix recognition errors. The fused hypotheses are passed to the LLM, yielding improved responses. 
\end{abstract}

\section{Introduction}
\label{sec:intro}
Large Language Models (LLMs) have significantly advanced the development of Automatic Speech Recognition (ASR), achieving excellent performance in general domains. However, their progress is hindered by numerous real-world problems with the models, training methods, and data. To promote this research, the MLC-SLM Challenge and Workshop introduce a new multilingual conversational speech dataset\footnote{https://www.nexdata.ai/competition/mlc-slm}, \footnote{https://github.com/mubingshen/MLC-SLM-Baseline} to support research in multilingual speech language modeling.

The challenge consists of two tracks, each designed to evaluate distinct aspects of speech language modeling. Task I focuses on multilingual ASR, where systems are provided with oracle speaker labels and segmentation boundaries and are evaluated using Word Error Rate (WER) or Character Error Rate (CER). Task II, in contrast, addresses joint speaker diarization and recognition without oracle information and is evaluated using Diarization Error Rate (DER) and tcpWER/tcpCER.

In this paper, we focus on Task I. Experiments show that keyword spotting is severely disturbed in noisy environments. Inspired by the End-to-End contextual biasing approach\cite{Shakeel_2024}, we deploy an ASR system designed for keyword correction.

\begin{figure}[ht]
  \centering
  \begin{subfigure}[t]{0.48\textwidth}
    \centering
    \includegraphics[width=\textwidth]{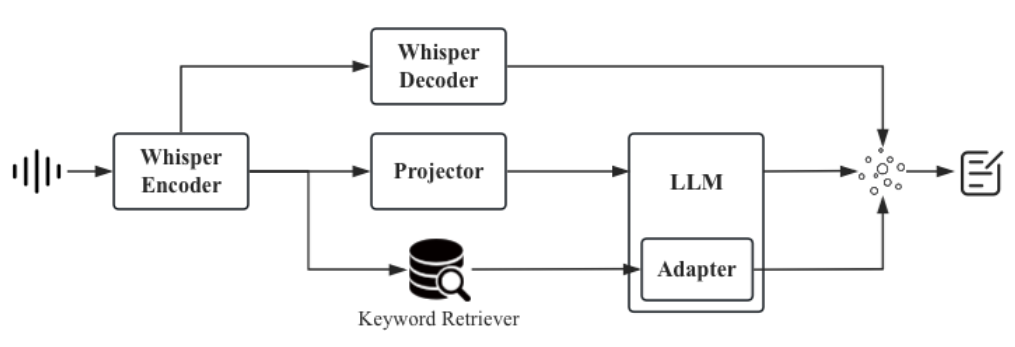}
    \caption{Structure overview of our model.}
    \label{fig:encoder}
  \end{subfigure}\\
  \hfill
  \begin{subfigure}[t]{0.48\textwidth}
    \centering
    \includegraphics[width=\textwidth]{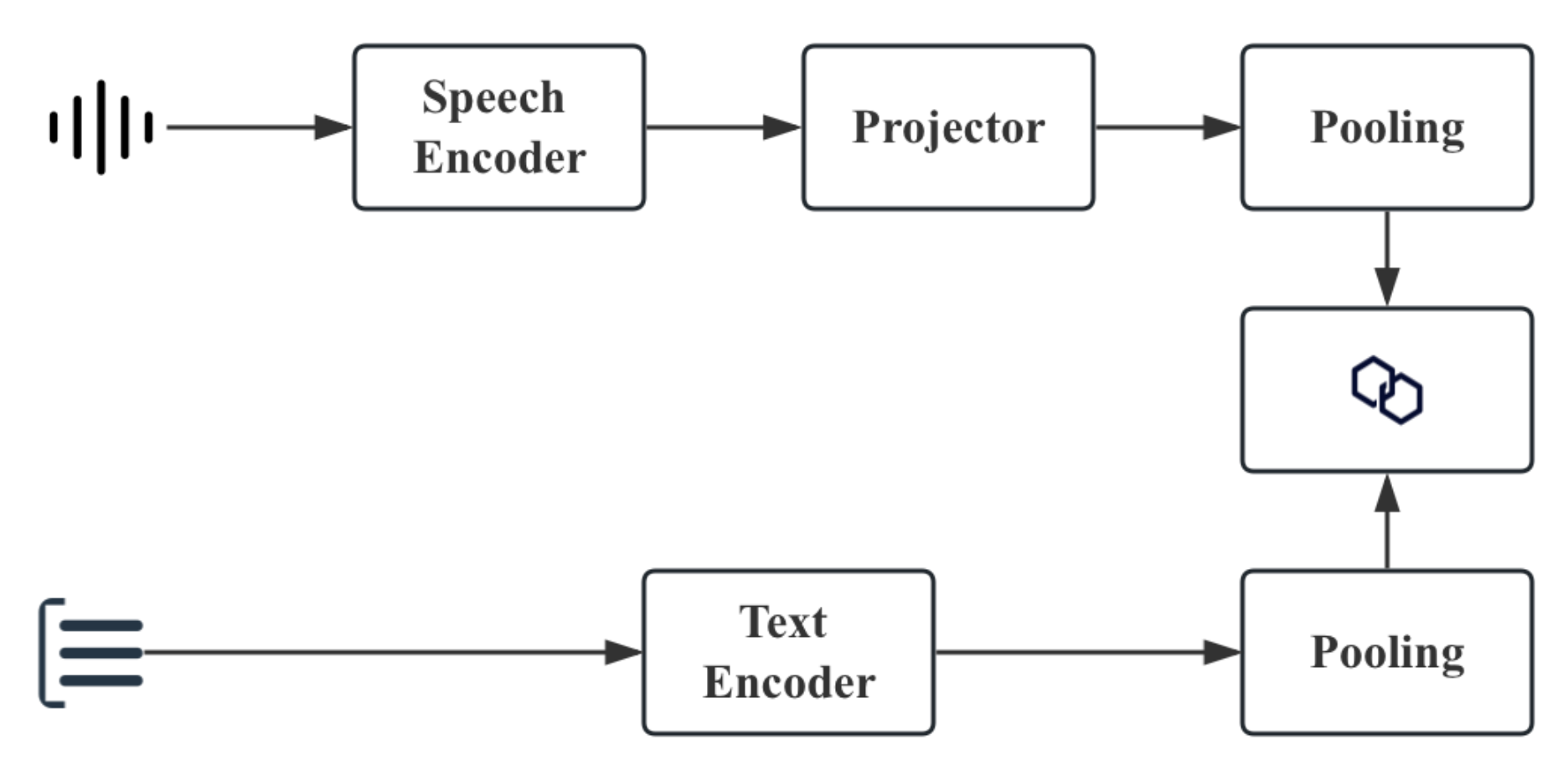}
    \caption{Keyword retrieval via cross-modal similarity.}
    \label{fig:retriever}
  \end{subfigure}
  \caption{Structural overview of the proposed ASR pipeline and the design of the keyword retriever, which incorporates Attention Pooling as the pooling module.}
  \label{fig:overview}
\end{figure}

\section{Proposed Method}
\label{sec:proposed}
As shown in Figure \ref{fig:overview}, our system consists of four key components: Speech Encoder, LLM, Projector, and Keywords Retriever. 

\subsection{Basic Architecture}
Our ASR system skeleton adopts the basic SLAM-ASR\cite{ma2024embarrassinglysimpleapproachllm} architecture, which consists of an audio encoder, a projector, and an LLM decoder. The projector is based on convolution layers and MLP to achieve alignment of speech embedding to LLM embedding.

\subsection{Keywords Retriever}
We trained the keyword retriever using a CLAP-style\cite{elizalde2022claplearningaudioconcepts} contrastive learning approach to identify possible keywords directly from speech. As a cross-modal keyword retriever, we adopted the classic dual-tower model, as shown in Figure \ref{fig:retriever}. We reused existing model components to simplify the system architecture. Specifically, the speech processing tower utilized the Whisper\cite{radford2023whisper} model and its supporting Projector in the SLAM architecture, whereas the text processing tower directly employed the LLM itself. Then, through a pooling layer, we projected the features of these two modalities into a unified shared embedding space. To implement the LLM generation process to optimize the information from keywords, we fine-tune an LLM Adapter using LoRA\cite{hu2021loralowrankadaptationlarge} technique.

\subsection{Result Aggregation}
We use a weighted fusion module to integrate three intermediate recognition outputs into the final transcription, namely the output from the whisper decoder, the output from the original LLM, and the RAG-based output. Note that the output based on RAG is given a higher weight.

\section{Experiments}
\label{sec:exp}
\subsection{Datasets and Metrics}
We use the English (American) dataset of MLC-SLM to train and evaluate (Development Set) the model. This subset comprises individuals of diverse genders and ages, with a natural conversation style. The word error rate of the label is lower than 2\%.

The performance of the model is evaluated primarily by the word error rate (WER), which serves as the core evaluation indicator. 

In addition, we introduce a Semantic Consistency Metric(SEM) to evaluate the impact of recognition errors on the meaning of the sentence. This metric is useful for downstream tasks that rely on ASR qualities, such as QA tasks. We used a pre-trained sentence transformers/all-roberta-large-v1\footnote{huggingface.co/sentence-transformers/all-roberta-large-v1} model, which is good at extracting semantic representations of sentences, to calculate the cosine distance as the measurement.

\subsection{Model Configuration}
A pretrained openai/whisper-large-v3-turbo\footnote{huggingface.co/openai/whisper-large-v3-turbo} is selected as the speech encoder. Whisper is known for its strong generalizability and robust speech representation. We only use its encoder to extract high-dimensional acoustic and semantic features. For LLM, we use the pre-trained Phi-4-mini-Instruct\footnote{huggingface.co/microsoft/Phi-4-mini-instruct}, which excels in the following instruction and text generation. The LLM has only 3.8B parameters, which reduces the system deployment overhead.

The projector, which plays a bridge role in connecting embedding spaces, is trained using the AdamW optimizer with the following parameters: an initial learning rate of 3e-4 and a weight decay of 0.01. The learning rate scheduler uses the Cosine decay strategy with a warm-up ratio of 0.01. Thanks to the compactness of the Phi-4-mini-instruct model, the training process can be completed on a single NVIDIA RTX 3090.

Once the SLAM-ASR model has been optimized, we train the retriever using contrastive learning with one positive and ten randomly selected negative keywords per sample. The retriever shares the same projector structure with SLAM-ASR, and its pooling layer is jointly trained to fit the retrieval task better. The keywords' embeddings we obtained are stored by FAISS\cite{johnson2017billionscalesimilaritysearchgpus} for efficient retrieval. Last, we manually selected 2,000 high-quality keyword correction samples and fine-tuned LLM for 5 epochs, achieving better error correction capabilities.

\subsection{Results}

\begin{table}[htbp]\footnotesize
\centering
\caption{Experiment Results (the Keywords RAG methods are based on Phi-4-mini-Instruct and Whisper Large-v3-turbo)}
\begin{tabular}{llll}
\hline
                                                                                                 &                          & WER   & SEM    \\ \hline
\multicolumn{1}{l|}{\multirow{3}{*}{\begin{tabular}[c]{@{}l@{}}Challenge\\ Baseline\end{tabular}}}                                                   & Vanilla Whisper-large-v3 & 14.14 & /      \\
\multicolumn{1}{l|}{}                                                                            & Baseline-Qwen            & 13.83 & /      \\
\multicolumn{1}{l|}{}                                                                            & Baseline-Llama           & 16.97 & /      \\ \hline
\multicolumn{1}{l|}{\multirow{2}{*}{\begin{tabular}[c]{@{}l@{}}Whisper\end{tabular}}} & tiny                     & 33.66 & 0.8217 \\
\multicolumn{1}{l|}{}                                                                            & large-v3-turbo           & 26.67 & 0.8940 \\ \hline
\multicolumn{1}{l|}{\multirow{3}{*}{SLAM-ASR}}                                                   & phi+tiny                 & 29.55 & 0.7480 \\
\multicolumn{1}{l|}{}                                                                            & phi+large-v3-turbo       & 15.09 & 0.8762 \\
\multicolumn{1}{l|}{}                                                                            & qwen+large-v3-turbo      & 15.31 & 0.8673 \\ \hline
\multicolumn{1}{l|}{\multirow{2}{*}{Keywords RAG}}                                               & w/o FT                   & 32.98 & 0.8085 \\
\multicolumn{1}{l|}{}                                                                            & w FT                     & 16.06 & 0.8726 \\ \hline
\multicolumn{1}{l|}{Proposed Method}                                                             & RAG-Boost                        & 11.67 & 0.9132\\ \hline
\end{tabular}
\label{tab:res}
\end{table}

As shown in Table \ref{tab:res}, our proposed method achieves consistently better results in both recognition accuracy and semantic relevance. In particular, it outperforms the baseline in terms of the word error rate (WER), indicating improved robustness and transcription quality. Overall observations are as follows:\\
\begin{enumerate}
\item The encoder strength matters most. Replacing Whisper-tiny with phi or large-v3-turbo consistently yields the most significant single jump in WER.

\item Context helps, but only after co-training. Raw RAG degrades performance; guided fine-tuning (ours, Keyword-RAG w FT) is essential for harnessing external knowledge.

\item Semantic robustness (SEM) tracks but lags in WER. Systems that slash WER (SLAM, ours) can still lose a few SEM points unless semantic objectives are explicitly optimized; our method recovers both.

\item The remaining gap (around 12\% WER) is dominated by overlapped speech and overly rapid speech per manual inspection. Future work could explore language modeling conditioned on the speaker role or lattice-aware retrieval to push below 10 \% WER.
\end{enumerate}

In addition, we compare the performance of Phi-4-mini-Instruct and Qwen2-7B-Instruct\footnote{huggingface.co/Qwen/Qwen2-7B-Instruct} under the same training conditions. The results show that Phi-4 outperforms Qwen, while it has only 3.8B parameters. Such results highlight its superior adaptability to ASR tasks. This further reveals the promising potential of compact language models in speech recognition applications.\\

\section{Conclusions}
\label{sec:conclude}

In this paper, we propose RAG-Boost (ST-ShinozakiLab Task I system), which enhances the baseline LLM-based ASR system of the MLC-SLM Challenge (task I) with a retrieval-augmented generation (RAG) module on the fly. The proposed approach achieves high performance in WER and high semantic matching, validating the design choice of jointly aligned speech-to-text retrieval training on noisy conversation data.

\bibliographystyle{IEEEtran}\footnotesize
\bibliography{reference}

\end{document}